# Using CNNs for AD classification based on spatial correlation of BOLD signals during the observation

Nazanin Beheshti[1]   Lennart Johnsson[1]

**Abstract**

Resting state functional magnetic resonance images (fMRI) are commonly used for classification of patients as having Alzheimer's disease (AD), mild cognitive impairment (MCI) or being cognitive normal (CN). Most methods use time-series correlation of voxel signals during the observation period as a basis for the classification. In this paper we show that using Convolutional Neural Network (CNN) for classification based on spatial correlation of time-averaged signals yield a classification accuracy of up to 82% (sensitivity 86%, specificity 80%) for a data set with 429 subjects (246 cognitive normal and 183 Alzheimer patients). For the spatial correlation of time-averaged signal values we use voxel subdomains around the center points of the 90 regions AAL atlas. We form the subdomains as sets of voxels along a Hilbert curve of a bounding box in which the brain is embedded with the AAL regions center points serving as subdomain seeds. The matrix resulting from the spatial correlations of the 90 arrays formed by the subdomain segments of the Hilbert curve yields a symmetric 90x90 matrix that is used for the classification based on two different CNN networks, a 4-layer CNN network with 3x3 filters and with 4, 8, 16 and 32 output channels respectively, and a 2-layer CNN network with 3x3 filters and with 4, 8 output channels respectively. The results for the two networks are reported and compared.

## 1. Introduction

Alzheimer's disease is progressively degenerative with a character of memory loss, mood and behavior changes, and deepening confusion about time and place. Alzheimer's disease is thought to begin 20 years or more before symptoms arise (Beason-Held et al. 2013) with small changes in the brain that are unnoticeable to the person affected. The cause of the disease is that neurons involved in thinking, learning, and memory have been damaged or destroyed (Mohs and Davis 1986) schematically illustrated in figure 1. A healthy adult brain has about 100 billion neurons (Herculano-Houzel 2009) each with long branching extension called synapsis with a healthy brain having about 100 trillion synapsis (Zimmer 2011). They transmit information in tiny bursts of chemicals that are released by one neuron and detected by receiving neurons. The time scale of these communication processes ranges from microseconds to seconds (Rudas et al. 2020). It is estimated that worldwide about 50 million people are affected by AD, but that only about 25% of those affected have been diagnosed with AD. The lifetime per patient care cost of AD is estimated to about$250k (Association 2019). and the total cost of care of AD patients could exceed $1 trillion by 2050.

Functional magnetic resonance imaging (fMRI) (Ogawa et al. 1990) is a non-invasive tool to study the function of the brain. In fMRI activity in the brain is observed over time with a spatial (voxel) resolution typically of 2 – 4 mm and a sampling rate of 0,5 – 2 Hz (Ogawa et al. 1990) for a duration of a few minutes. The brain activity is captured from Blood Oxygen Level-Dependent (BOLD) magnetization (Buxton 2013) detected by the MRI scanner (Ogawa et al. 1990). For the study of AD versus healthy subjects Resting-State fMRI (RS-fMRI) (Biswal et al. 1995)(Beckmann et al. 2005) is commonly used, i.e., BOLD signals are captured when subjects are resting. A major emphasis in the field is on the analysis of resting state functional connectivity that is measured by the time-series correlation of BOLD signals for a pair of voxels, or more often correlation of BOLD signal time series formed by averaging the time-series for collections of voxels in Regions of Interest (ROI). Degradation in resting state functional connectivity have been identified in disorders like Alzheimer's (Rombouts et al. 2005) (Khosla et al. 2019) (Damoiseaux 2012) disease.

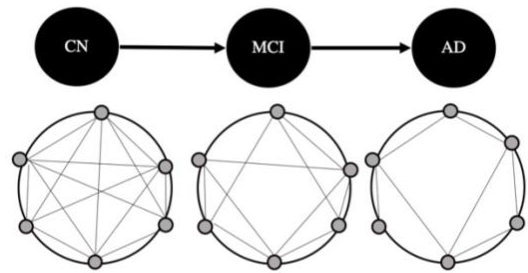

*Figure 1*: Resting State Network for Cognitive Normal (CN), Mild Cognitive Impairment (MCI), and Alzheimer's disease (AD)

---

[1] Department of Computer Science, University of Houston

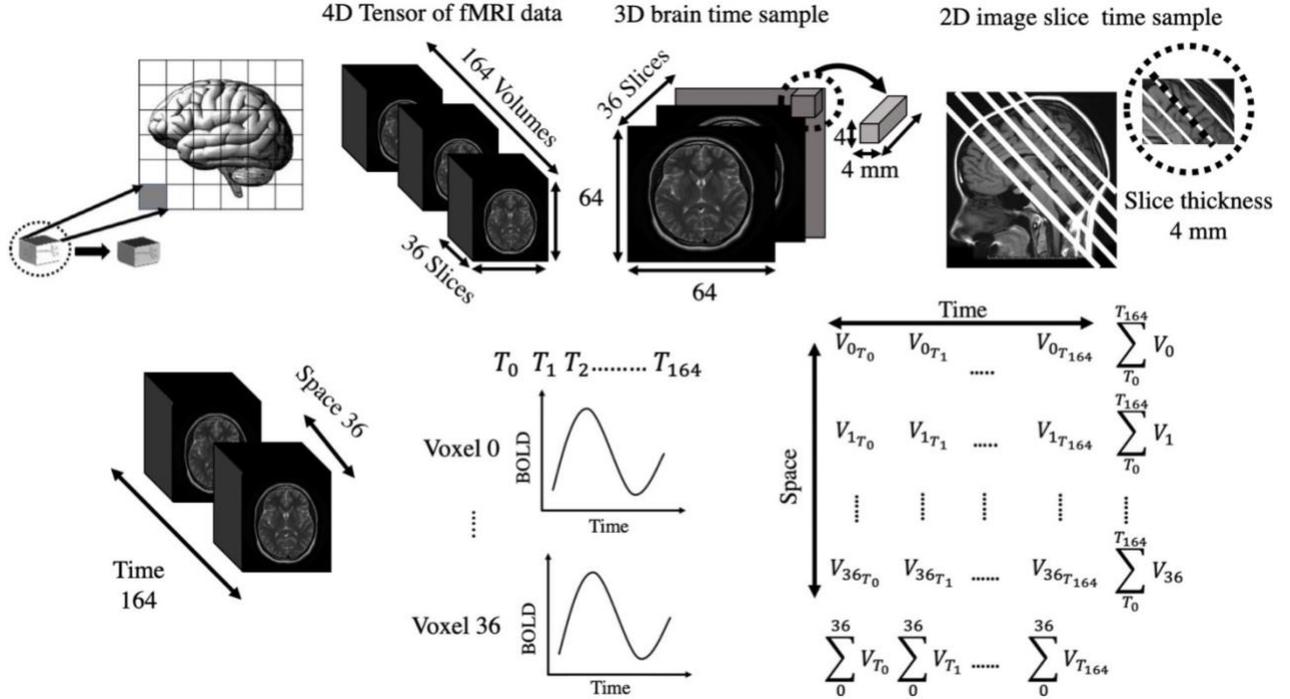

*Figure 2:* 4D fMRI data in time, Functional Magnetic Resonance Images consist of 164 volumes in time, each volume consists of 36 slices, and each slice is a matrix of 64x64. fMRI record Blood Oxygen Level of blood (BOLD) for a duration of time. For every region, time series signals of all voxels inside the region which have similar dynamic can be summed up or for each voxel, time series signal of the voxel is averaged.

Our approach to the classification of three groups of subjects, Cognitive Normal (CN), Alzheimer's Disease (AD), and Mild Cognitive Impairment (MCI) subjects, is to use spatial correlations between ROIs of time-averaged voxel BOLD signals. For the correlation, an array is formed for each ROI from the time-averaged voxel BOLD signals along a segment of a Hilbert curve traversing a bounding box enclosing the brain with segments centered at the AAL-90 region atlas (Lancaster et al 2000) center points. We evaluate different sizes of the ROIs by using different lengths Hilbert curve segments. The spatial correlations of the ROI Hilbert curve segments form a symmetric 90x90 matrix that is used as input for a convolutional neural network (CNN). We studied the effectiveness of a 4-layer CNN with 3x3 filters and 4,8,16 and 32 output channels respectively, and a 2-layer CNN using 3x3 filters and 4 and 8 output channels respectively.

**Contributions:**
1. Forming ROIs from segements of a Hilbert curve traversal of a bounding box enclosing the whole brain and assessing the functional connectivity homogeneiry of these ROIs.
2. Classification based on spatial correlation between ROIs of time-averaged voxel BOLD signals.
3. Use of convoltional neural networks on the ROI correlation matrices for the classfication task.

## 2. Related work

With voxels in fMRI for AD classification having spatial extent in the 2 – 4 mm range an anatomical brain region contains hundreds to thousands of voxels (Bowman, Guo, and Derado 2007)(Wu et al. 2013) with voxels containing a few hundred thousand neurons. With anatomical brain regions containing many voxels clustering of voxels into ROIs is often used to reduce computational complexity in functional studies. The BOLD signal for an ROI is typically determined as the spatial average of the ROI voxel BOLD signals. Clustering of voxels into ROIs may also improve signal to noise when the BOLD signals of the voxels within an ROI are highly correlated while the noise is uncorrelated. The "seed" voxel for an ROI can be chosen based on the strength of the BOLD signal, as the center point of an anatomical region, as in our case, or some measure of centrality (Bonacich 1987). The ROI can then be formed around a seed based on spatial proximity (Damoiseaux 2012), as in our case, or the strength of the functional connectivity with some cut-off criteria. The latter generally results in regions of arbitrary shape and size and may not be spatially localized, whereas the former may result in ROIs with a highly non-uniform functional connectivity within the ROI (Yu et al. 2012). Some studies have shown that coherence of ROIs can affect clustering accuracy (Craddock et al. 2012).

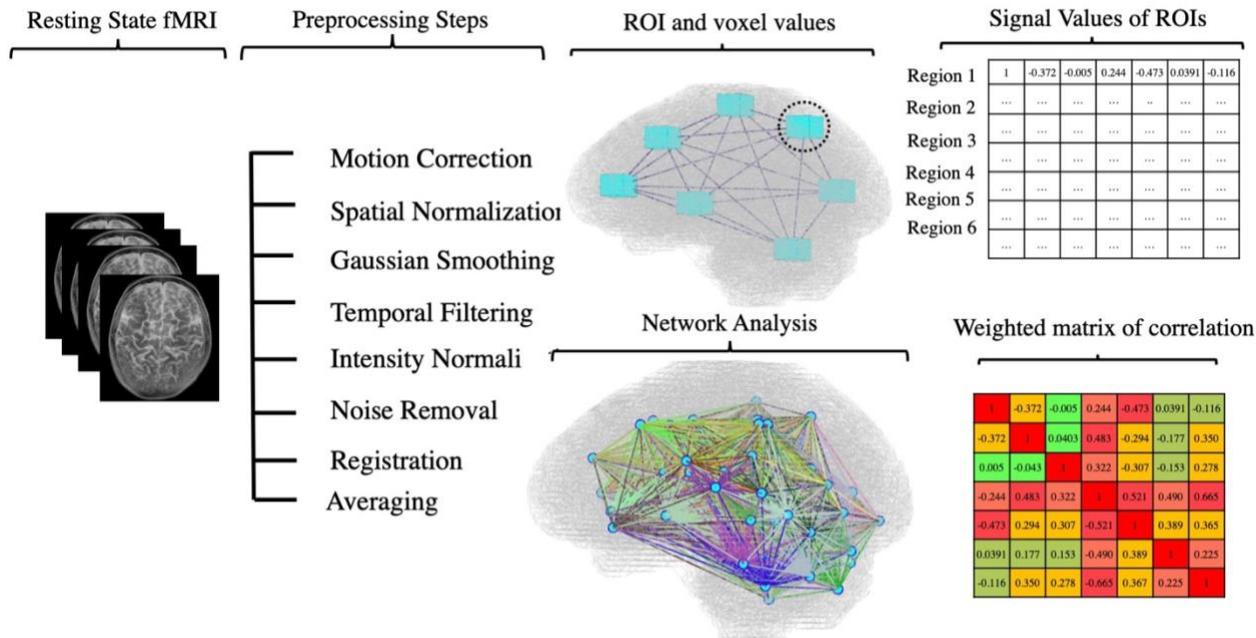

*Figure 3*: fMRI classification pipeline, Preprocessing steps in fMRI volumes and slices, grouping voxels around the seed voxels and generating regions of interest, taking signals of voxels in every region and find the correlation between every pair of regions. The nodes of the graph are regions and the edges are correlation values between regions.

Methods not based on spatial proximity such as principal component analysis (PCA) (Pearson 1901) and independent component analysis (ICA) (Lee 1998) have been used for reducing the computational effort in classification. Graph theoretical approaches have also been applied to the brain functional connectivity to characterize communication in the brain, using measures such as clustering coefficients, node degree, betweenness, path lengths, local efficiency, global efficiency and modularity for classification (Yu et al. 2012)(Xiang et al. 2020)(Khazaee, Ebrahimzadeh, and Babajani-Feremi 2015) (Chen et al. 2011)(Jie et al. 2013)(Challis et al. 2015). In some studies, extracted graph measures are used as discriminative features for a machine learning approach like an SVM classifier (Fornito, Zalesky, and Breakspear 2013).

## 3. Data Sets

For our assessment of the effectiveness of spatial correlation of time-averaged BOLD signals and CNNs for classification of AD vs MCI and CN subjects we created a RS-fMRI data set with 526 subjects of which 183 were diagnosed with AD, 97 with MCI, and 246 were CN, see Table 1. The OASIS dataset was acquired using single-shot gradient echo planar imaging (EPI) with total of 164 whole brain time samples using a repetition time (TR) of 2.2s (total duration 360 seconds), an echo time (TE) of 27ms, a flip angle (FA) of 90, and slice resolution 64 x 64 4 x 4 mm voxels of 4 mm thickness with 36 slices to cover the brain volume. The ADNI dataset was acquired using Gradient echo (GR) pulses, with different range from 46 to 200 time samples, a repetition time (TR) of 3s (total duration 420s), an echo time (TE) of 30ms, a flip angle (FA) of 80, and a slice resolution of 64 x 64 3.3 x 3.3 mm voxels of 3.3 mm thickness with 48 slices to cover the brain volume. Both RS-fMRI data sets were acquired using 3 Tesla Siemens scanners.

*Table 1:* summarize socidempgraphic information of subjects under study

| OASSIS DataSet | | | | |
|---|---|---|---|---|
| | CN | AD | MCI | Total |
| Number | 139 | 85 | 0 | 224 |
| Male/Female | 46/92 | 38/46 | 0 | |
| Age | 64±9.17 | 74±8.10 | 0 | |
| ADNI DataSet | | | | |
| Number | 107 | 98 | 97 | Total |
| Male/Female | 42/65 | 35/63 | 45/52 | 302 |
| Age | 76±7.19 | 75±7.69 | 74±8.6 | |

## 4. Data Preprocessing

The BOLD signals for the whole brain are acquired as a sequence of 2D adjacent slices (36 and 48 respectively for the data sets we used) as shown in figure 2. Hence, BOLD signals for different slices are collected at slightly different times. In the pre-processing the BOLD signals are transformed to a common reference time for each whole brain scan by simply adjusting the time $t_k$ for slice k by (N/2+1-k) x TR/N, with k=1,…N where N is the number of slices and TR is the time for a whole brain scan, thus shifting slice sampling time to that of the middle slice.

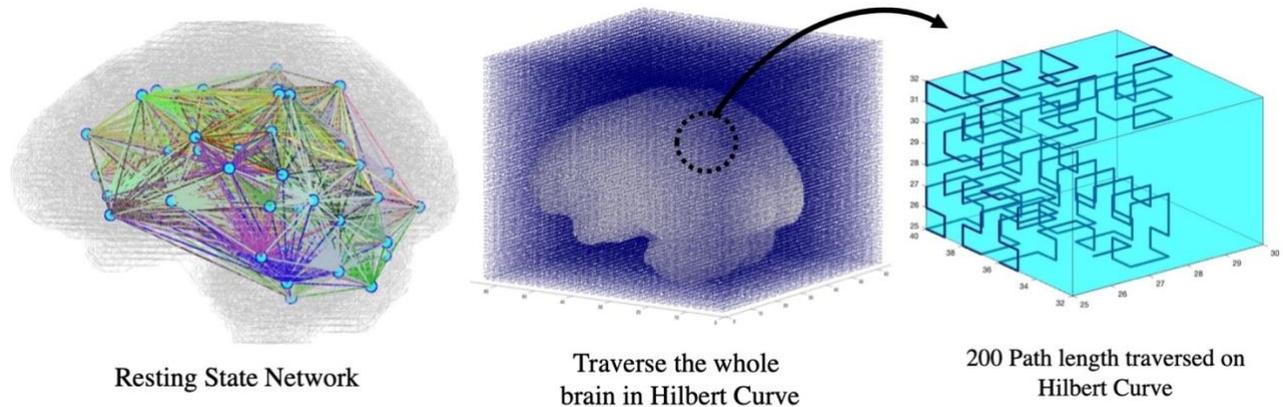

*Figure 4*: Filling 3D brain space with Hilbert Curve to preserve spatial locality between voxels, take the seed voxels based on AAL atlas map and group the voxels around the seed voxel on the Hilbert Curve with different path length.

For RS-fMRI data from different subjects to be comparable, the collected data must be mapped to a common reference brain. In this work we use the MNI-152 (Grabner et al. 2006) brain template as the reference brain to which individual subject brain images are mapped using affine transformations (rotation, translation, shear, scaling) by first aligning the structural (MRI) and functional (fMRI) images for each subject, called co-registration then map the co-registered fMRI images to the MNI-152 brain known as normalization. For co-registration, we use the SPM12-V7771 software package (Ashburner et al. 2014) that use an entropy and normalized mutual information based objective function, equation (1) (Studholme, Hill, and Hawkes 1999). In equation (1) $I(X,Y)$ is the mutual information between X and Y and $H(.)$ denotes entropy. For the alignment, misalignment tolerances are given for translation, rotation, scaling, and shear which in our case are (0.02, 0.001, 0.01, 0.001)

$$NMI = \frac{I(X,Y)}{\sqrt{H(X)H(Y)}} \qquad (1)$$

The normalization uses a tissue probability map containing prior probabilities of all the tissues found in the image. The prior probabilities of different tissue classes at each location in the brain are constructed from a large number of brains mapped to a reference brain, MNI-152 in our case. For normalization a log-likelihood criteria (Ashburner and Friston 2005) has proved effective. We use the software package SPM12 also for the normalization. For the MNI-152 reference brain we use 3x3x3 mm voxels thus requiring a 3D interpolation of the BOLD signals from the 4x4x4 mm OASIS and the 3.3x3.3x3.3 mm ADNI voxel used in data acquisition. After time alignment, co-registration, normalization and signal interpolation for voxel size, spatial smoothing of BOLD signals was made using a Gaussian filter with the Full Width at Half Maximum (FWHM) of 8 mm to enhance the signal to noise ratio.

After preprocessing, the bounding box for the OASIS and ADNI data has 53x63x52 3x3x3 mm voxels, 173,628 in total. To put this voxel count in perspective, the brain volume on average is 1130 cm$^3$ for women and 1260 cm$^3$ for men. Hence, with a 27 mm$^3$ voxel volume, about 41,851 and 46,660 voxels, respectively, encompasses the actual brain volumes. Thus, of the 173,628 voxels of the bounding box for our data set brains only occupy about 25%. Our bounding box used for the Hilbert curve is of shape 64x64x64 for ease of Hilbert curve construction and hence has 262,144 voxels resulting in only about 1/6$^{th}$ of the Hilbert curve voxels being inside the brain volumes. By selecting the ROI seed voxels as AAL-90 region center points and limiting the Hilbert curve segments to extend at most 100 voxels along the curve away from the center points, the risk of a ROI extending beyond the brain is low. To asses how the time-averaged BOLD signal varies across each ROI we compute the average over all subjects *i* of the time-averaged BOLD signal for each region $j$ and voxel $k$ in a region, $VI^{SA}_{j,k}$ as given in equation (2) where $VI_{i,j,k}$ is the time-averaged BOLD signal for subject $i$, region $j$ and voxel $k$. $VI^{SA}_{(j,k)}$ the time-averagd BOLD signal for all subjects, regions, and voxels has mean and standar deviation of 12,692 , and 2,155. We also compute the average time-averaged BOLD signal for all subjects and each ROI seed voxels, $SI^{SA}_j$ equation (3) where $SI_{i,j}$ is the time-avergaed BOLD signal over all subjects for the ROI seed voxel for subject $i$, and region $j$. $SI_{(i,j)}$ has mean and standard deviation 13,588, and 1,887 respectively. Figure 7 shows histograms for the time-averaged BOLD signal values in all 90 ROIs. 71% of voxels have intensities in the range of 10,000 to 14,000 which is close to the 13,588 average seed voxel intensity.

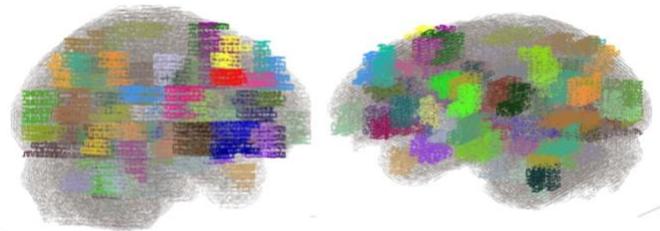

*Figure 5:* color-coded Hilbert curve segments plots inside the brain area after traversing brain voxels

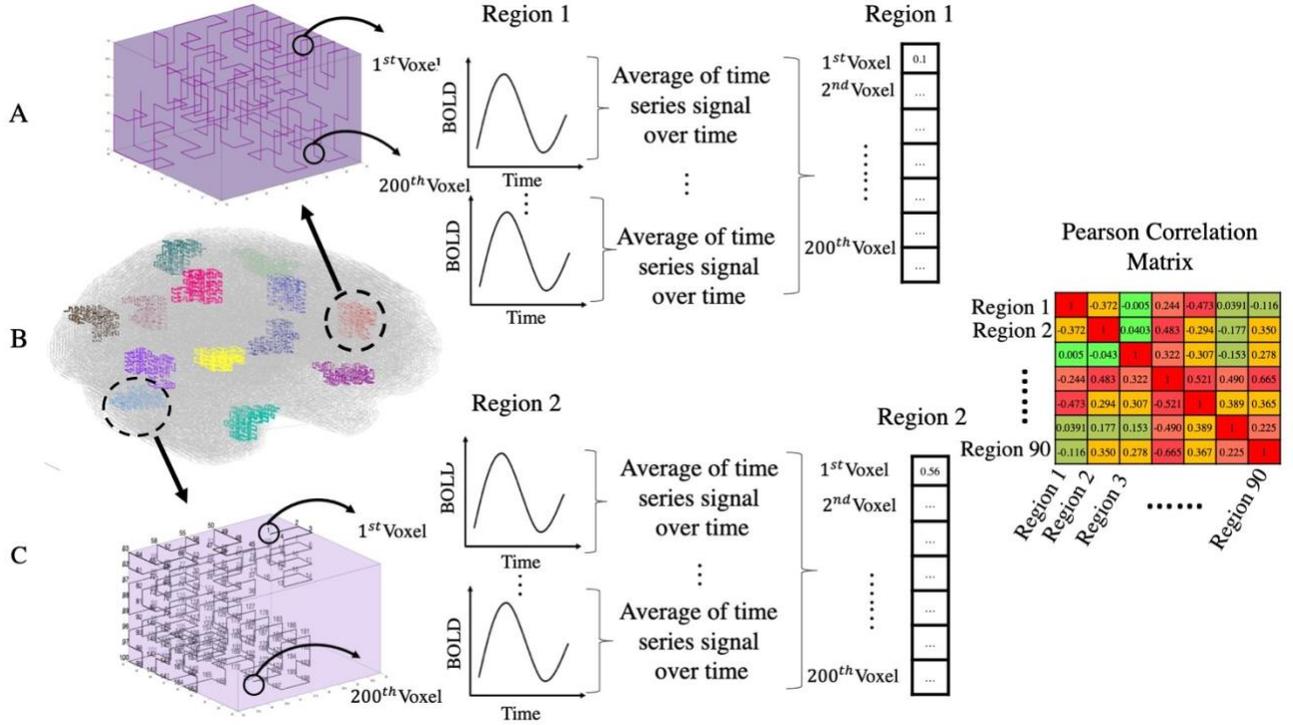

*Figure 6*: taking average of time series signal over time for all traversed voxels along the path on Hilbert Curve and taking the Pearson correlation between every pair of 90 regions

$$VI^{SA}_{(j,k)} = \frac{1}{526}\sum_{i=1}^{526} VI_{i,j,k} \qquad (2)$$

$$SI^{SA}_{(j)} = \frac{1}{526}\sum_{i=1}^{526} SI_{(i,j)} \qquad (3)$$

## 5. Methodology

### 5.1 Overview

Our goal is to design a computationally and energy efficient CNN for accurate AD, MCI, and CN classification. As basis for the classification we use spatial correlation of ROIs formed from segments of a Hilbert curve traversal of a bounding box enclosing the OASIS and ADNI brain data sets. as shown in figures 4, 5 and 6. We use one ROI for each of the AAL-90 region (Lancaster et al. 2000) with the region center point serving as ROI seed voxel. The extent of a ROI is determined by Hilbert curve segments symmetric around the seed voxel. To assure ROIs are distinct we limit the segment length to at most 201 voxels, since for larger segment lengths some segments overlapped. Time-average voxel BOLD signal values are used to represent the BOLD signal for a voxel. Two different traversal paths for two different regions are shown in figure 6. The voxel BOLD signal values for a ROI form an array of Hilbert curve ordered values.

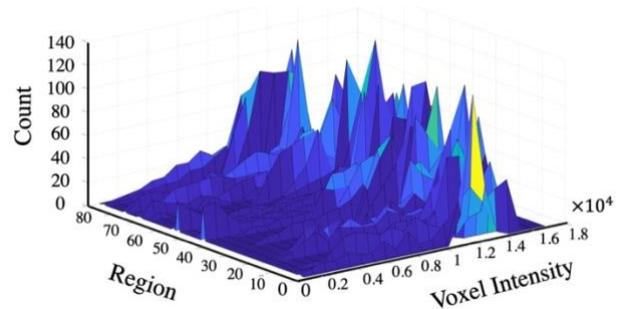

*Figure 7*: Histogram of Voxel intensities along the path on the segments Hilbert Curve for 90 ROIs averaged over all subjects

Table 2: Average (%) and standard deviation (%) of TN, TP, FP, FN on thirty different test sets for four experiments

| Network | Path | CN vs AD (57% vs 42%) (320 Train 109 Test) | | | | CN vs AD (52% vs 47%) (164 Train 41 Test) | | | |
|---|---|---|---|---|---|---|---|---|---|
| | | TN | TP | FP | FN | TN | TP | FP | FN |
| 4,8,16,32 | 201 | **85±5** | **79±6** | **14±5** | **20±6** | 77±9 | 78±9.2 | 22±9.2 | 21±9.2 |
| | 101 | 83±4.78 | 70±5.7 | 16±4.7 | 29±5.7 | 92±6 | 82±10 | 8±6 | 17±10 |
| 4,8 | 201 | 84±5.8 | 77±5.9 | 15±5.8 | 22±5.9 | 87±7.7 | 86±7.3 | 12±7.7 | 13±7.3 |
| | 101 | 85±4.4 | 74±5.3 | 14±4.4 | 25±5.3 | **91±6.9** | **86±9.4** | **8.7±6.9** | **13±9.4** |

| Network | Path | MCI vs CN (47% vs 52%) (164 Train 40 Test) | | | | MCI vs AD (49% vs 50%) (156 Train 39 Test) | | | |
|---|---|---|---|---|---|---|---|---|---|
| | | TN | TP | FP | FN | TN | TP | FP | FN |
| 4,8,16,32 | 201 | 79±10 | 72±22 | 20±10 | 18±8.6 | 80±7.6 | 81±7.8 | 20±7.6 | 19±7.8 |
| | 101 | 90±5.9 | 92±6.2 | 9.3±5.9 | 7.4±6.2 | 85±9.6 | 80±7.6 | 15±9.6 | 20±7.6 |
| 4,8 | 201 | 85±9.4 | 83±8.9 | 14±9.4 | 16±8.9 | 84±5.9 | 84±7.8 | 16±5.9 | 16±7.8 |
| | 101 | **91±6.1** | **92±6.9** | **8.4±6.1** | **7.7±6.9** | 90±8.9 | 80±8.1 | 10±8.9 | 20±8.1 |

These arrays are used for pairwise spatial Pearson correlation (Benesty et al. 2009) of ROIs and result in a symmetric matrix of correlation values which is the basis for the CNN. The Pearson correlation $\rho(V, W)$ between two regions is defined in equation (4), in which N is the ROI Hilbert curve segment length (101 or 201 in this study) and $V$ and $W$ are arrays of time-averaged voxel BOLD signals traversed along the Hilbert curve segments for ROI $V$ and $W$. $\mu_V, \mu_W$ are the mean of the voxel values of regions $V$ and $W$, respectively. $\sigma_V$ and $\sigma_W$ are the variances of values of $V$ and $W$. $\rho(V, W) = \frac{1}{N-1}\sum_{i=1}^{N}(\frac{\overline{V_i}-\mu_V}{\sigma_V})(\frac{\overline{W_i}-\mu_W}{\sigma_W})$ (4)

### 5.2 Regional Functional Homogeneity

As a measure of functional homogeneity (ReHo) of voxel BOLD signals in an ROI we compute the Pearson time-series correlation of voxel BOLD signals between all pairs of voxels $k$ and $z$ in the ROI $j$ for each subject $i$, $PC_{(i,j,k,z)}$ in equation (5). $BOLD_{i,j,k,t_m}$ is the time series BOLD signal for subject $i$, region $j$, voxel $k$, at time $t_m$, m=1, 2,.M with a total of M time samples. $BOLD_{i,j,k}^A$ is the time-averaged BOLD signal in subject $i$, region $j$, voxel $k$. The correlation coefficients range between -1 and +1. If two signals have strong time correlation the coefficient is close to +1.

$$PC_{(i,j,k,z)} = \frac{\sum_{1}^{M}(BOLD_{i,j,k,t}-BOLD_{i,j,k}^A)(BOLD_{i,j,z,t}-BOLD_{i,j,z}^A)}{\sqrt{\sum_{1}^{M}(BOLD_{i,j,k,t}-BOLD_{i,j,k}^A)^2 \sum_{1}^{M}(BOLD_{i,j,z,t}-BOLD_{i,j,z}^A)^2}}$$ (5)

$$PC_{i,j}^A = ReHo_{(i,j)} = \frac{\sum_{k=1}^{Region\ Size}\sum_{z=1}^{Region\ Size}(PC_{(i,j,k,z)})}{Region\ Size \times Region\ Size}$$ (6)

$$ReHo_i^{A1} = \frac{1}{90}\sum_{j=1}^{90}ReHo_{(i,j)}$$ (7)

$$ReHo_j^{A2} = \frac{1}{224}\sum_{i=1}^{224}ReHo_{(i,j)}$$ (8)

$$ReHo_i^{std1} = \frac{1}{90}\sqrt{\sum_{j=1}^{90}(ReHo_{(i,j)}-ReHo_i^{A1})^2}$$ (9)

$$ReHo_j^{std2} = \frac{1}{224}\sqrt{\sum_{i=1}^{224}(ReHo_{(i,j)}-ReHo_j^{A2})^2}$$ (10)

As a measure of the ROI homogeneity, ReHo$_{i,j}$ for subject $i$ and ROI $j$ we use the average of the Person time-correlation

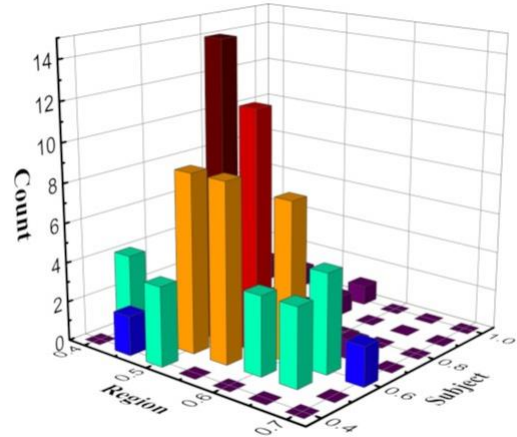

Figure 8: (a) represent the distribution of functional homogeneities within subjects, (b) represent the distribution of functional homogeneities within regions

for the region as shown in equation (6). We assessed the average and standard deviation of the functional homogeneity of the ROIs for each of the 224 subjects of the OASIS dataset, equations (7,9) and for each ROI for all subjects, equations (8,10). The OASIS data set is relatively new compared to the ADNI data set for which several functional homogeneity studies have been published. Figure 8 represents the distribution of ReHo values.

Table 3: Average (%) and standard deviation (%) of classification accuracy on thirty different test sets for four experiments.

| Network | Path | CN-AD (320 Train 109Test) | | | CN-AD (164 Train 41 Test) | | | MCI-CN (164 Train 40Test) | | | MCI-AD (156Train 39Test) | | |
|---|---|---|---|---|---|---|---|---|---|---|---|---|---|
| | | ACC | SE | SP | ACC | SE | SP | ACC | SE | SP | ACC | SE | SP |
| 4,8,16,32 | 201 | **82±3.4** | **86±5.1** | **80±6.2** | 78±5.5 | 78±9 | 79±9.2 | 80±5.5 | 79±10.5 | 81±8.6 | 80±4.8 | 80±7.6 | 81±7.8 |
| | 101 | 78±4 | 84±4.7 | 71±5.7 | 86±5.5 | 92±6 | 82±10 | 91±4.2 | 90±5.9 | 92±6.2 | 82±5 | 85±10 | 80±8 |
| 4,8 | 201 | 81±3.3 | 85±5.8 | 77±5.9 | 86±5.4 | 87±7.7 | 86±7.3 | 84±5.3 | 85±9.4 | 83±8.9 | 83±4.8 | 84±5.9 | 84+7.8 |
| | 101 | 80±3.4 | 85±4.4 | 75±3.3 | **89±4.9** | **91±6.9** | **87±9.4** | **91±4.4** | **91±6.1** | **92±6.9** | **84 ± 6** | **90±9** | **80±8** |

### 5.3 Neural Network Architecture

The input to the CNNs we study is the 90x90 ROI spatial correlation matrices for each subject with one channel. The number of output channels for the 4-layer CNN are 4,8,16,32 respectively, with each having 3x3 filters and the last layer with 32 channels being a fully connected layer. The 2-layer CNN has 4 and 8 output channels with the last layer also being fully connected. The total number of parameters for a convolution layer using $K \times K$ filters having $C_i$ input channels and $C_o$ output channels is $K \times K \times C_i \times C_o$. Thus, the first layer the 4-layer CNN with 1 input channel and 4 output channels has 36 parameters, the 2nd layer has 288 parameters, and the 3rd layer has 1,152 parameters. The fourth fully connected layer with 32 output channels has 16x12x12x32=73,728 parameters. The total number of parameters is 75,204 (36+288+1,152+73,728). With single precision data representation, the 4-layer CNN model size is about 293.77 kiB and the 2-layer CNN has 3×3×4+45×45×4×8=64,836 parameters. The 2-layer network model size is about 253 kiB. which is small compared to many other CNN models. The proposed 4-layer network is trained in 94.36 seconds with 320 training 90x90 correlation matrices using Tensorflow.1.14, Python 3.7.3 on a 2.3GHz Intel Core I5-7360U processor with 2 cores and 4 threads, 16 GB 2133 MHz LPDDR3 memory, and macOS Mojave-10.14.6.

### 5.4 Training

To evaluate the classification accuracy achieved by our CNNs, four experiments were carried out. In all four experiments, the dataset is divided into training and testing based on the 80/20 Pareto Principle (Sanders 1987). We report the average and standard deviation of True Positives (TP), True Negative (TN), False Positive (FP), False Negative (FN), Specificity TN/(TN+FP), Sensitivity TP/(TP+FN) and Accuracy (TP+TN)/(TP+TN+FP+FN). In the first experiment (AD-CN) the 97 MCI subjects were excluded from our data set resulting in a set with 429 subjects. Of the 109 (20%) subjects were uniformly randomly chosen for the test set and the remaining 320 were used for training. Training sets that had a larger than 20% difference between AD and CN subjects were discarded. Thirty training/test sets with less than 20% difference between AD and CN subjects were generated. In the second experiment, (AD-CN) 164 subjects are used for training and 41 subjects are used for test set. In the third experiment, (MCI-CN), 164 subjects are used for training and 40 subjects are used for test set. In the fourth experiment (MCI-AD), 156 subjects are used for training and 39 subjects are used for test set.

For experiments two - four 30 training and test sets were generated as described for experiment one. No data augmentation was made in any of the experiments. Adam Optimizer (Kingma and Ba 2014) with a fixed learning rate of 0.0001 was used. On every epoch, one batch of 4 matrices is picked with batches picked in a way ensuring every matrix is used only once. In each epoch, we shuffle the training set, so in every epoch a different batch of matrices are input to

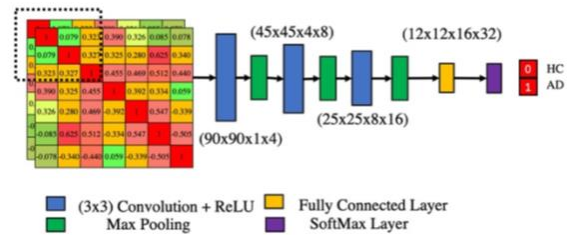

Figure 9: Four Layer neural Network

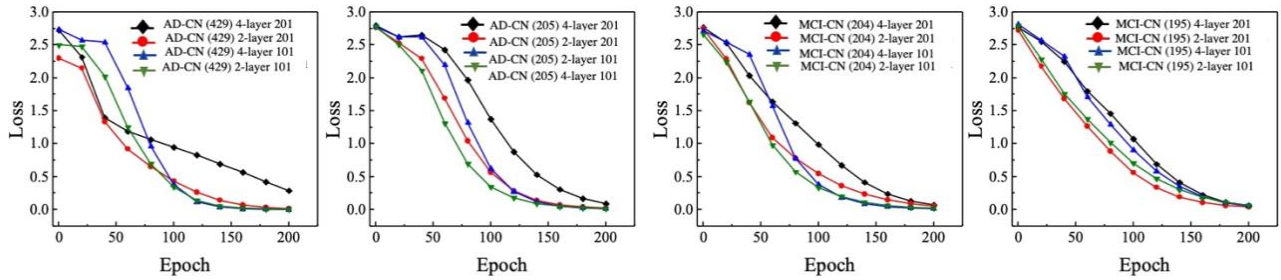

*Figure 10*: Training Loss sampled at interval of 25 epochs in the range of 0 to 200 is represented for four experiments with two different path on segments of Hilbert Curve (101,201 including seed voxel) and for 4-Layers, and 2-Layers networks.

## 6. Experimental Results

To investigate the effect of ROI size on classification ROIs with 101 and 201voxels were evaluated. As is shown in table 3, the highest accuracy in classification of CN vs AD with 320 subjects in the training set and 109 in the test set was achieved by the 201 ROI size and the 4-layer CNN. For the CN vs AD classification with a 164 subject training set and 41 subject tests set the highest accuracy was achieved for the 101 ROI size and the 2-layer CNN. The 101 ROI size and the 2-layer CNN also yielded the best accuracy for the MCI vs CN and MCI vs AD classifications. All layer-2 trainings converged between 100 to 200 epochs. The training of the 4-layer network in the first experiment converged in 150 to 200 epochs.

## 7. Conclusion

The 2-layer CNN with only about 20% of the parameters of the 4-layer CNN resulted in higher accuracy, sensitivity and specificity than the 4-layer network for the three experiments with about 200 subjects with subjects split about 80%20% between training and tests with no subjects in common between training and test sets. For these test sets the 101 ROI size also performed better than the 201 ROI size. For the experiment with 429 subjects and 201 ROI size the 4-layer CNN performed marginally better than the 2-layer CNN with respect to average accuracy, sensitivity, specificity: 82%, 86%, 80% (4-layer) vs 81%, 85%, 77% (2-layer). However, for the 101 ROI size the 2-layer CNN performed better with average accuracy, sensitivity and specificity being: 78%, 84%, 71% (4-layer) vs. 80%, 85%, 75% (2-layer). The reduction in average accuracy, sensitivity and specificity for the 2-layer network in using the 101 ROI size was less than for the 4-layer network for the 429 subject experiment and the 2-layer network performing almost as well for the 101 ROI size as the 4-layer network for the 201 ROI size.

*Table 4*: Distribution of Test and Training set in each experiment

| Test Set | | | Training Set | | |
|---|---|---|---|---|---|
| Min | Max | Avg | Min | Max | Avg |
| AD-CN (183, 246) | | | | | |
| AD | | | | | |
| 34% | 51% | 42% | 39% | 43% | 42% |
| CN | | | | | |
| 48% | 65% | 57% | 54% | 60% | 57% |
| AD-CN (98, 107) | | | | | |
| AD | | | | | |
| 34% | 65% | 48% | 43% | 51% | 57% |
| CN | | | | | |
| 34% | 65% | 51% | 48% | 65% | 52% |
| MCI-CN (97,107) | | | | | |
| MCI | | | | | |
| 32.5% | 62.5% | 46.4% | 43% | 51% | 47% |
| CN | | | | | |
| 37.5% | 67.5% | 53.5% | 48% | 56% | 52% |
| MCI-AD (97,98) | | | | | |
| MCI | | | | | |
| 35% | 66% | 48% | 45% | 53% | 49% |
| AD | | | | | |
| 33% | 64% | 51% | 46% | 54% | 50% |

**Acknowledgement**
302 samples of our data for this project was funded by the Alzheimer's Disease Neuroimaging Initiative (ADNI) (National Institutes of Health Grant U01 AG024904) and DOD ADNI (Department of Defense award number W81XWH-12-2-0012). ADNI is funded by the National Institute on Aging, the National Institute of Biomedical Imaging and Bioengineering, and through generous contributions from the following: AbbVie, Alzheimer's Association; Alzheimer's Drug Discovery Foundation; Araclon Biotech; BioClinica, Inc.; Biogen; Bristol-Myers Squibb Company; CereSpir, Inc.; Cogstate; Eisai Inc.; Elan Pharmaceuticals, Inc.; Eli Lilly and Company; EuroImmun; F. Hoffmann-La Roche Ltd and its affiliated company Genentech, Inc.; Fujirebio; GE Healthcare; IXICO Ltd.; Janssen Alzheimer Immunotherapy Research & Development, LLC.; Johnson & Johnson Pharmaceutical

Research & Development LLC.; Lumosity; Lundbeck; Merck & Co., Inc.; Meso Scale Diagnostics, LLC.; NeuroRx Research; Neurotrack Technologies; Novartis Pharmaceuticals Corporation; Pfizer Inc.; Piramal Imaging; Servier; Takeda Pharmaceutical Company; and Transition Therapeutics. The Canadian Institutes of Health Research is providing funds to support ADNI clinical sites in Canada. Private sector contributions are facilitated by the Foundation for the National Institutes of Health (www.fnih.org). The grantee organization is the Northern California Institute for Research and Education, and the study is coordinated by the Alzheimer's Therapeutic Research Institute at the University of Southern California. ADNI data are disseminated by the Laboratory for Neuro Imaging at the University of Southern California.

224 of our data were provided by OASIS, OASIS-3: Principal Investigators: T. Benzinger, D. Marcus, J. Morris; NIH P50 AG00561, P30 NS09857781, P01 AG026276, P01 AG003991, R01 AG043434, UL1 TR000448, R01 EB009352. AV-45 doses were provided by Avid Radiopharmaceuticals, a wholly owned subsidiary of Eli Lilly.